\pdfoutput=1

\documentclass[final,1p,times]{elsarticle} 
\usepackage{graphicx} 
\usepackage{amssymb} 
\usepackage{amsthm} 

\journal{Nuclear Physics A} 
\begin{document} 

\begin{frontmatter} 


\title{Open Heavy Flavor Production in Heavy Ion Collisions}

\author{J.C.~Dunlop}

\address[a]{Brookhaven National Laboratory, 
Upton, NY, 11973, USA}

\begin{abstract} 
The interaction of heavy partons, charm and beauty, with the matter 
created in heavy ion collisions has been of great interest in recent years.
Heavy partons were predicted to interact less strongly with the matter than light partons.  In apparent contrast to these predictions, unexpectedly strong suppression of non-photonic electrons from heavy flavor decays has been seen.  However, significant experimental uncertainties remain, both in the measurements themselves and in the separation
of the contribution from charm and beauty, which have 
complicated the interpretation of these results.  
The current experimental
situation is critically reviewed and 
prospects for making these measurements more easily 
interpretable discussed.  

\end{abstract} 

\end{frontmatter} 



\section{Introduction}
The mass of heavy partons, charm and beauty, introduces an
additional knob with which to adjust the interaction of 
partons with matter.  Partons moving slowly have been 
predicted to lose less energy in matter via QCD Bremsstrahlung,
 much as particles moving at low $\beta \gamma$ lose little
energy from ordinary QED Bremsstrahlung.
Original predictions of dramatic differences of the energy
loss of charm, and especially beauty, from that of light quarks ~\cite{Dokshitzer:2001zm,Zhang:2003wk} have not been seen experimentally.
Instead, in central Au+Au collisions at $\sqrt{s_{NN}} = $ 200 GeV, the spectra of non-photonic electrons, primarily from the
decay of hadrons containing charm and beauty, are suppressed nearly as much 
as light hadrons such as $\pi^{0}$~\cite{Abelev:2006db,Adare:2006nq}.
``Non-photonic'' electrons are those electrons from which contributions
from decays involving photons, predominantly $\pi^{0}$ Dalitz and 
conversions in detector material of the $\gamma$ daughters of $\pi^{0} \rightarrow \gamma \gamma$, have been subtracted.
This unexpectedly large suppression has led to further investigations into the mechanism of QCD energy
loss~\cite{Mustafa:2004dr,Djordjevic:2005db,Armesto:2005mz,Wicks:2005gt,vanHees:2005wb}, which continue to this day and to this conference.  At the current time,
it is a fair statement that the data are not fully understood theoretically, 
and do not have sufficient power to distinguish theoretical scenarios.  

In the coming decade, there will be a sea change in both the qualitative and quantitative 
constraints that experiments can place on these theoretical investigations.
Both STAR and PHENIX plan to install high precision vertex detectors
for separation of charm from beauty, and, as STAR's focus, for the direct reconstruction
of a number of charmed hadrons~\cite{Bouchet}.  Pb+Pb collisions will occur at the
Large Hadron Collider (LHC), at an order of magnitude higher collision energy
than is available at RHIC.  The LHC detectors are also tuned for the detection
of charm and beauty, with projection studies at ALICE very well developed~\cite{Dainese:2009ru}.
As a precursor for these studies, in the rest of these proceedings I 
will discuss critically the limitations of current measurements and
how they are expected to improve in the next few years.
 
\section{Total charm cross section}
\begin{figure}
\centering
\includegraphics[width=0.6\textwidth]{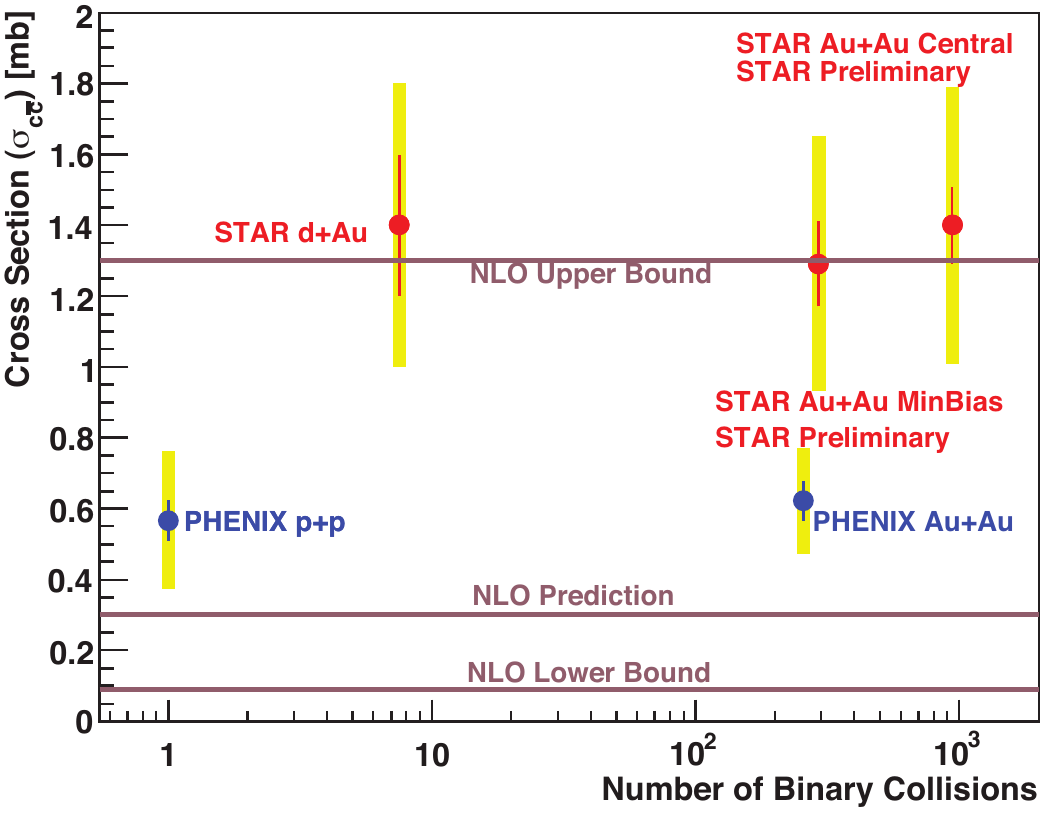}
\caption{
Total charm cross section as measured by STAR~\protect\cite{Adams:2004fc,Abelev:2008hja}
and PHENIX~\protect\cite{Adare:2006nq,Adler:2004ta}.
Calculation from~\protect\cite{Vogt:2007aw}.}
\label{fig:totalcross}
\end{figure}
The total charm cross section has been extracted by both 
STAR and PHENIX at $\sqrt{s_{NN}} = $ 200 GeV, and is compared to next to leading order theoretical calculations~\cite{Cacciari:2005rk,Vogt:2007aw} in 
Figure~\ref{fig:totalcross}. 
In $p+p$ collisions PHENIX has extracted this cross section
from the spectra of non-photonic electrons~\cite{Adare:2006hc}, and also from di-electron measurements~\cite {Adare:2008asa}.   STAR has extracted the 
$p+p$-equivalent charm cross section in d+Au collisions, assuming binary
scaling, from a combination
of reconstructed $D^0$ mesons and non-photonic electrons
~\cite{Adams:2004fc}.  Both PHENIX~\cite{Adler:2004ta} 
and STAR~\cite{Abelev:2008hja} have made equivalent measurements in  
Au+Au collisions, where the STAR measurements additionally contain
identified muons at extremely low $p_{T}$.   Further measurements
were shown at this conference.  Within STAR and PHENIX separately,
the total charm cross section is found to scale well with the number
of binary collisions, as expected if charm is produced in initial
hard scatterings in Au+Au collisions, but there is a discrepancy
of approximately a factor of 2 between the experiments.
Both sets of measurements are consistent with theoretical
calculations, within the large theoretical uncertainty.

\begin{figure}
\centering
\includegraphics[width=0.48\textwidth]{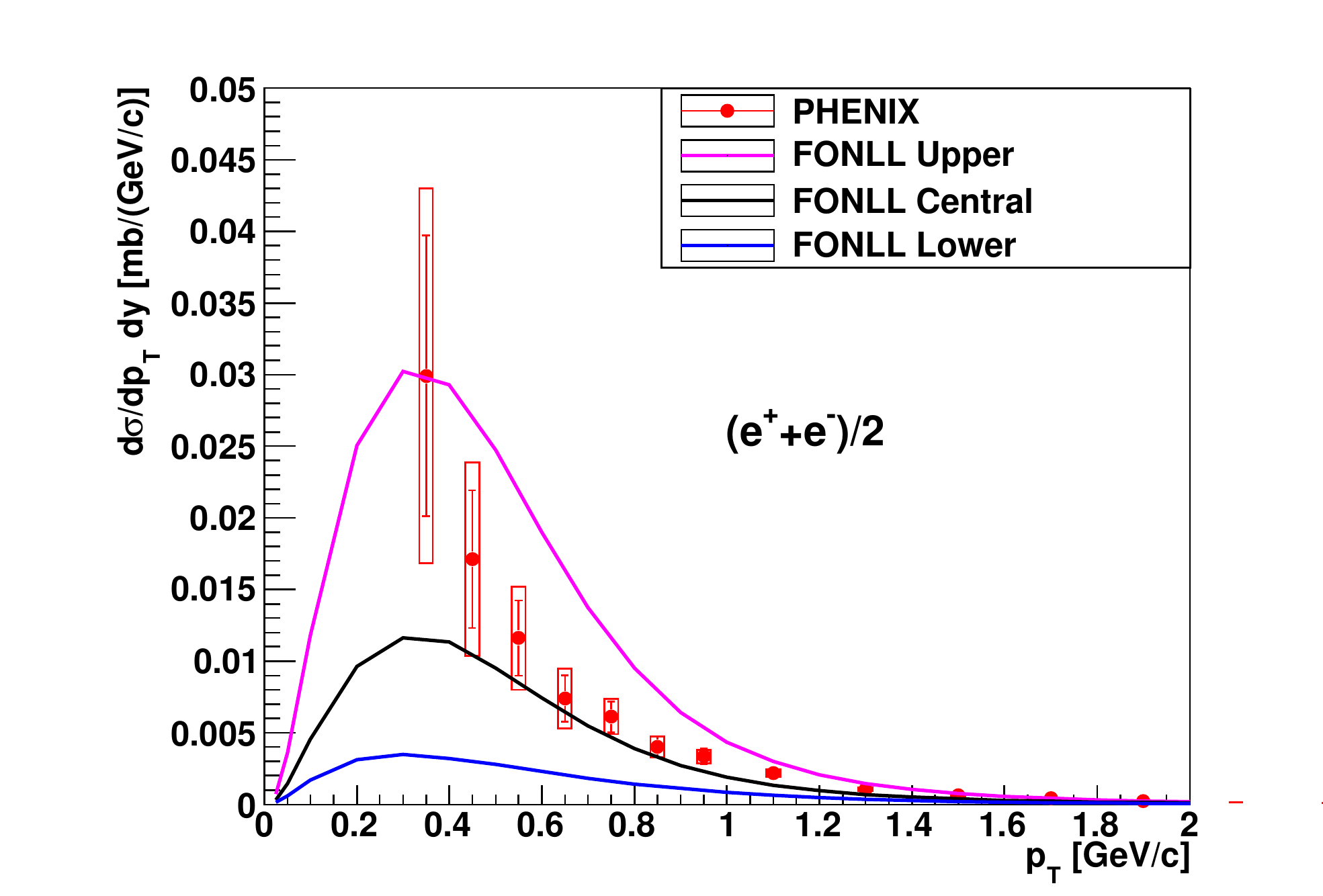}
\includegraphics[width=0.48\textwidth]{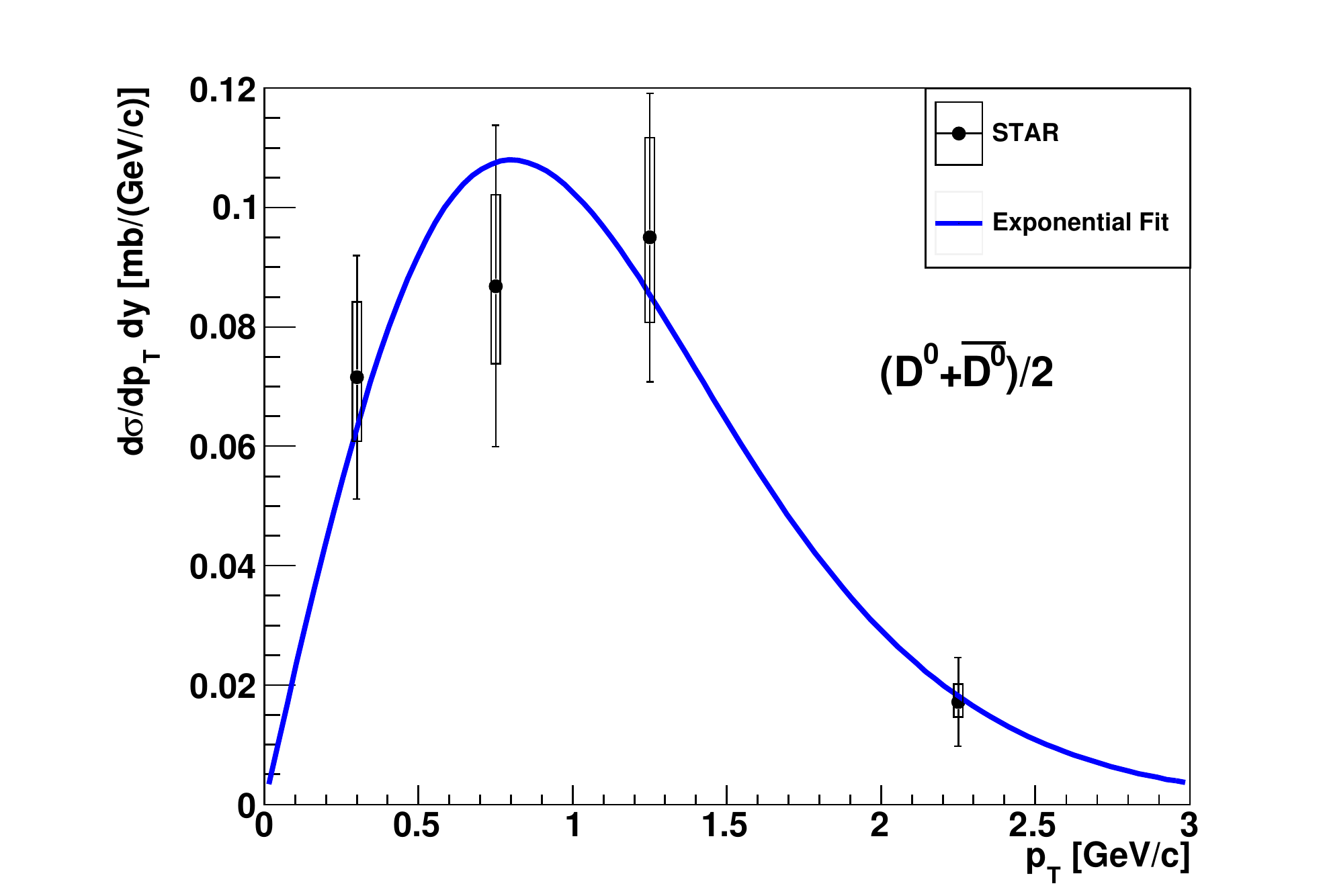}
\caption{Left: Differential cross section $d\sigma/dp_{T} dy$ for 
non-photonic electrons in $p+p$ collisions at $\sqrt{s}$ = 200 GeV.
Data is from~\cite{Adare:2006nq}, and calculations from ~\cite{Cacciari:2005rk}.
The data point for $p_{T}$ below 0.4 GeV/c is not used in the 
extraction of the total cross section.  Right: $p+p$-equivalent differential
cross section $(\sigma^{pp}_{inel}/N_{bin}) dN/(dy dp_{T})$ for $\left(D^0+\overline{D^0}\right)/2$ from d+Au collisions.  Line is
an exponential fit from the publication.  Data and fit from \protect\cite{Adams:2004fc}.
}
\label{fig:linear}
\end{figure}

The total charm cross section is a crucial input to models 
that incorporate regeneration of $J/\Psi$ in QCD matter
via coalescence.  In order to test these models, the charm
cross section needs to be known with precision.
Theoretical guidance is lacking, since as shown in Figure~\ref{fig:totalcross} the
 uncertainty in the 
calculated cross section ranges over nearly an order of magnitude. 
Both experimental measurements are lacking in precision.
The extraction of the total cross section by PHENIX suffers
from extrapolation beyond the measured range in $p_{T}$:
as shown in Figure~\ref{fig:linear} 
the measured single electron spectra are extrapolated by a factor of 1.8 to 
lower $p_{T}$ using the shape of the FONLL spectra, while
the di-electron yields are extrapolated to full phase space using PYTHIA.
The STAR electron measurements are even less constraining, since
their reach to low $p_{T}$ is smaller than that of PHENIX.
STAR extracted a total charm cross section from the $D^0$ alone,
and with a combined fit of $D^0$ and its electrons, and
found that the extracted total charm cross section changed by 
less than 10\%.   
As shown in Figure~\ref{fig:linear} the STAR $D^0$ 
measurements do not suffer from extrapolation issues at mid-rapidity, 
since they extend down to lower $p_{T}$ than the electrons and
the underlying spectrum is harder, since the parent $D$ meson is fully
reconstructed.  The $D^0$ measurements do 
suffer from large systematic uncertainties due
to large backgrounds and, with currently analyzed data, large statistical uncertainties.
In addition, assumptions need to be made about the fraction of charm
quarks that fragment specifically into $D^0$, 
and a large extrapolation by a factor of 4.7 is needed to convert the 
mid-rapidity $dN/dy$ yield to a total cross section.
No $D^{0}$ measurements have been made in $p+p$ collisions to date,
so binary scaling has to be assumed to obtain a $p+p$-equivalent cross section
from d+Au collisions.

To make progress towards specifically this observable, 
more precise measurements are needed.  For the total 
charm cross section, a reduction of extrapolation assumptions
is critical to make precision measurements. STAR plans
to measure $D^0$ mesons in $p+p$ collisions, 
with much higher statistics and additional background
rejection from its new Barrel Time of Flight detector in Run 9. 
STAR had a Silicon Vertex Tracker installed in the Au+Au run in 
Run 7, and at this conference presented first measurements from this run of
the $D^0$ meson using displaced vertex techniques
to reduce combinatorial background~\cite{LaPointe}.
In the future, background rejection will be greatly improved
by the higher precision Heavy Flavor Tracker, allowing precise
measurements of a number of states including the $\Lambda_{c}$, 
which will constrain the assumptions about the fraction
of charm quarks that fragment into $D^0$.
The extrapolation to full rapidity can be 
addressed with single muons in the PHENIX muon arms.
Current measurements are limited by systematics 
from backgrounds~\cite{Dion}, which are expected
to greatly decrease with the PHENIX FVTX upgrade.
Extrapolation to low $p_{T}$ will likely remain an issue
for such measurements. 

It is not clear, however, that the total cross section
provides fully relevant information.  The cross section
is dominated by low $p_{T}$: the mean $p_{T}$ of $D$ mesons
is approximately 1 GeV, and, after semi-leptonic decay,
approximately half of the non-photonic electrons from charm
lie at $p_{T}$ below 0.4 GeV, with the precise fraction depending
on theoretical input exactly where theoretical uncertainties are largest.
 Measurements of the total 
charm cross section therefore provide little constraint
for electrons, dileptons, or regenerated $J/\Psi$ at moderate $p_{T}$.
For dilepton studies in which charm is the background,
there is the additional issue of correlations between
charm quark and anti-quark pairs, and their possible modification in 
the heavy ion environment.  Such correlations will simply need to be measured
to make precise statements.  Displaced electron and muon pairs provide
one way to measure these correlations.  Electron-muon correlations can also
potentially help this situation, since they provide a rather specific measure of
correlated charm.  First proof-of-principle measurements
have been made~\cite{Engelmore}, but these clearly need further
luminosity, and without further detector
upgrades such as the STAR Muon Telescope Detector cannot be measured in the same phase space as di-electrons or
di-muons.
For partonic energy loss studies, the relevant $p_{T}$ range
is even higher, and behavior at high $p_{T}$ is largely independent of the 
behavior at low $p_{T}$ due to the steeply falling $p_{T}$ spectrum. 
\begin{figure}
\centering
\includegraphics[width=0.9\textwidth]{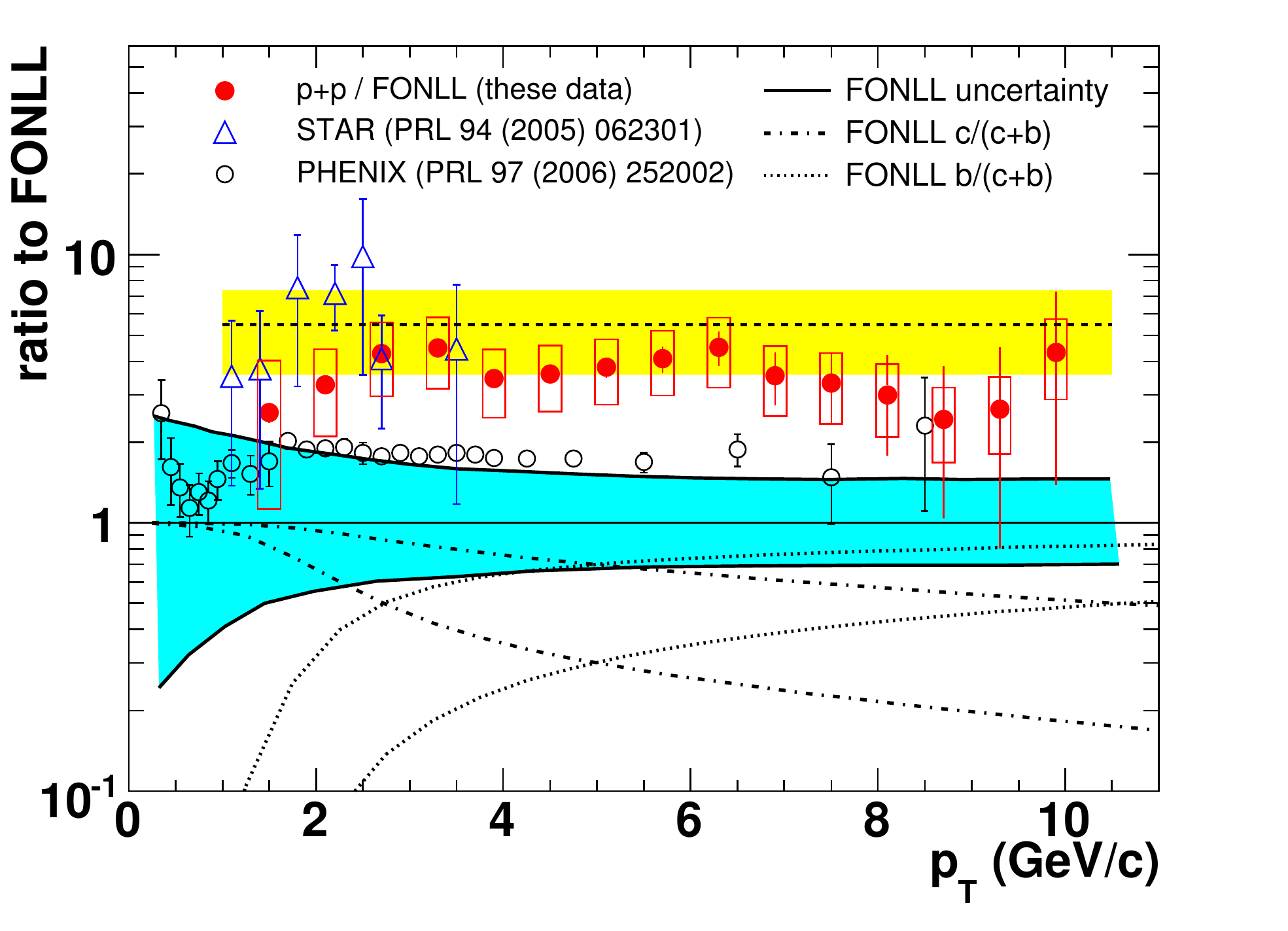}
\caption{Ratio to FONLL calculations of differential cross sections of
non-photonic electrons in proton-proton collisions.  Figure from~\protect\cite{Abelev:2006db}. 
}
\label{fig:FONLL}
\end{figure}

\section{Partonic energy loss}
Both STAR and PHENIX have measured non-photonic electrons in both
p+p and Au+Au collisions
to $p_{T}$ of approximately 10 GeV, as shown in Figure~\ref{fig:FONLL}.
As in the total cross section, the STAR measurements are approximately
a factor of 2 larger than those from PHENIX.  The reason for this discrepancy
is different, however.  The STAR total cross section is dominated by
$D^{0}$ measurements, but the comparison in Figure~\ref{fig:FONLL} is 
is between non-photonic electrons, 
two sets of measurements that should be directly comparable.
As with the total charm cross section, the binary-scaled ratio between Au+Au
and $p+p$ collisions, $R_{AA}$, is consistent between experiments.

One of the differences between STAR and PHENIX is the larger amount of material
in STAR prior to Run 8, due to the presence of the Silicon Vertex Tracker 
and Silicon Strip Detector.  This leads to a much larger background
of photonic electrons, from photon conversions, in STAR than in PHENIX.
This background is a major source of systematic error.  In order to reduce
this systematic error, in Run 8 the Silicon was removed from the interior
of the STAR detector.  This did have the expected effect of reducing conversion
backgrounds by approximately an order of magnitude, but final results were not ready
at the time of this conference.

PHENIX presented an investigation at this conference~\cite{Dion} in which
a further background was identified.  With the measured PHENIX
non-photonic electron and $J/\Psi$ differential cross sections,
electrons from $J/\Psi$ decays
are an appreciable contribution to the non-photonic electron
spectrum for $p_{T}$ above 5 GeV/c.  This contribution does not enter
into theoretical calculations that assume that the only source of non-photonic
electrons is decay of charm and beauty. The relative fraction
of this contribution depends upon the differential cross-sections
of the electrons themselves, and of the
$J/\Psi$ at high $p_{T}$.
The latter has been measured with 
some precision in Cu+Cu and $p+p$ collisions
by both STAR~\cite{Abelev:2009qa} and PHENIX~\cite{Adare:2006kf,Adare:2008sh}, 
but clearly remains statistics starved. 
$J/\Psi$ production at high $p_T$ has
not been measured with any precision in Au+Au collisions, and
so assumptions about possible $R_{AA}$ of $J/\Psi$
fold into systematics of the comparison of non-photonic electron $R_{AA}$ to
theoretical calculations.  In the PHENIX analysis, this contribution is small
but significant enough to be worth subtracting.

\section{Extraction of beauty}
Even once the non-photonic electron experimental discrepancies are resolved, and
non-photonic electrons measured with high precision, the question of charm
and beauty partonic energy loss will not be answered.  
Non-photonic electrons come from a mixture of the decays of charm and beauty,
the relative fraction of which has large theoretical uncertainties.  
Predictions of energy loss differ greatly between charm and beauty.
At $p_{T}$ of approximately 10 GeV/c, expectations
are that charm is relativistic enough to act essentially as a light quark, while beauty
is slow enough to remain less strongly interacting with the medium.
In order to test these predictions with certainty, beauty must be separated from charm.

\begin{figure}
\centering
\includegraphics[width=0.48\textwidth]{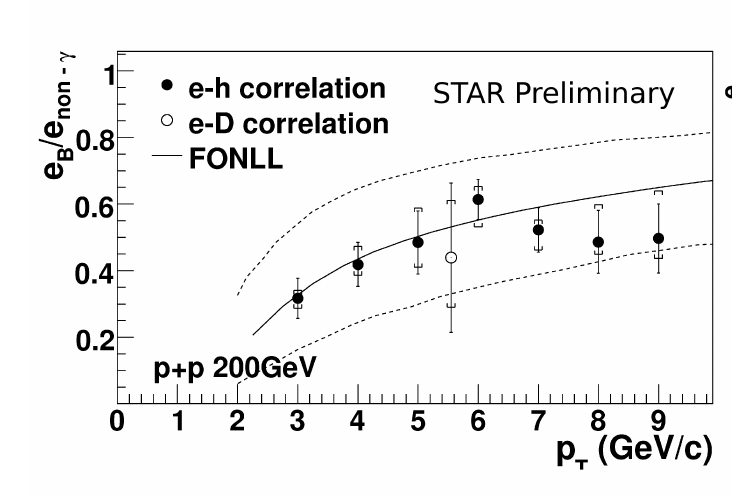}
\includegraphics[width=0.48\textwidth]{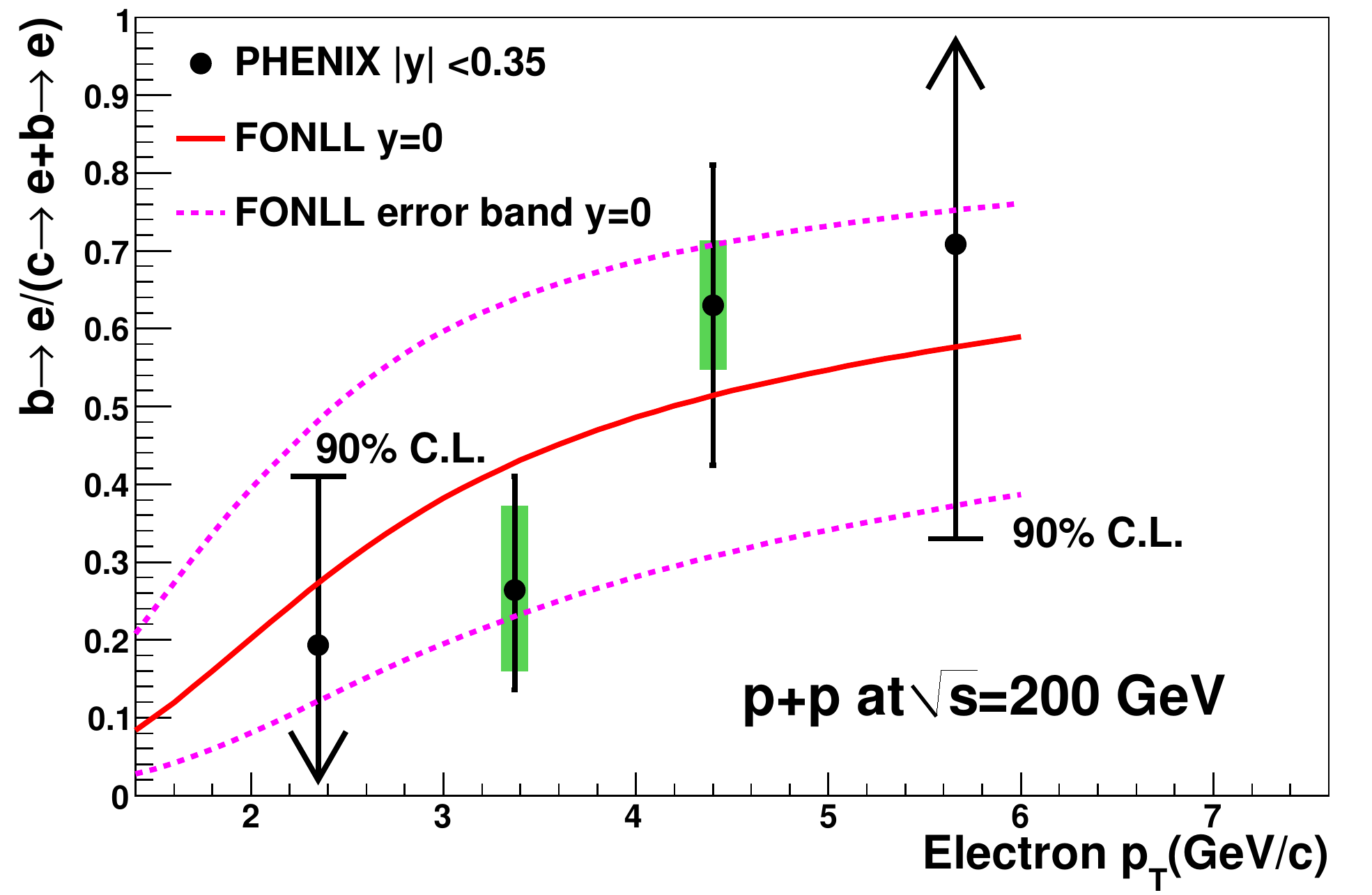}
\caption{Fraction of non-photonic electrons from beauty decays
in $p+p$ collisions, 
as compared to FONLL calculations.  Left: STAR~\protect\cite{Biritz}.  Right: PHENIX~\protect\cite{Adare:2009ic}.}
\label{fig:beautyfrac}
\end{figure}

As shown in Figure~\ref{fig:beautyfrac}, both STAR~\cite{Biritz} and PHENIX~\cite{Adare:2009ic} have attempted
to separate charm from beauty in $p+p$ collisions
utilizing electron-hadron correlations.
These techniques rely on the heavier mass of a $B$ meson than a $D$ meson, which
leads to clearly distinguishable patterns in the correlations of electrons
with hadrons.  STAR additionally has electron-$D^{0}$ correlations, with fully
reconstructed $D^{0}$ mesons, which decreases somewhat the dependence
on models but currently suffers from poorer statistics.
   
\begin{figure}
\centering
\includegraphics[width=0.7\textwidth]{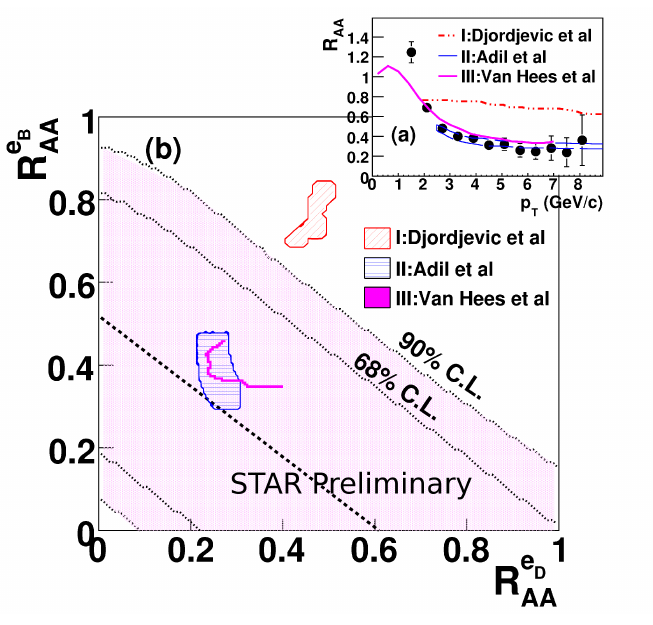}
\caption{Allowed region for $R_{AA}$ for electrons from beauty decays ($R^{e_{B}}_{AA}$)
vs. electrons from charm decays ($R^{e_{D}}_{AA}$), for electron $p_{T}\ >$ 5 GeV/c, from combination of the STAR
$R_{AA}$ for total non-photonic electrons, in the inset~\protect\cite{Abelev:2006db}, and the STAR beauty fraction
$e_{B}/(e_{B}+e_{D})$ from Figure~\protect\ref{fig:beautyfrac}.  Curves are theoretical
calculations from ~\protect\cite{Djordjevic:2005db,vanHees:2005wb,Adil:2006ra}, 
expressed as envelopes of the predictions across the $p_T$ range over which the figure integrates.  
}
\label{fig:beautyconstraint}
\end{figure}

These methods have not been successfully applied to Au+Au collisions to date.
Correlations have been measured~\cite{Biritz,Vale}, 
but statistical significance is low and 
interpretation is complicated by possible modification of the correlation
patterns.  To separately measure charm and beauty $R_{AA}$, there are 
therefore four unknowns and only three measurements (i.e. total non-photonic electron
spectra in $p+p$ collisions, total non-photonic electron spectra in Au+Au collisions,
and the fraction of non-photonic electrons that come from beauty decay in 
$p+p$ collisions).  With this information, STAR has
formed an exclusion region in $R_{AA}^{e_B}$ as a dependent variable vs.
 $R_{AA}^{e_D}$ as an independent variable in order to quantify the constraints
from this data.
This exclusion region is shown in Figure~\ref{fig:beautyconstraint}. 
  While it is physically reasonable to assume that the $R_{AA}$ for
charm is smaller than that for beauty, there is nothing in the
data to constrain this assumption, so the exclusion region contains all possible
$R_{AA}$ that would be consistent with the three sets of measurements.
This leads to a finite probability that the $R_{AA}$ for electrons from 
charm is greater than unity
and balanced by an extremely low $R_{AA}$ for electrons from beauty. 
Overall, the exclusion region appears 
to support finite suppression of beauty at 90\% Confidence Level.
The maximum $R_{AA}$ for electrons from beauty is approximately 0.9, at extremely
small charm $R_{AA}$, and for an $R_{AA}$ for electrons from charm of 0.2, as for the $\pi^{0}$, 
$R_{AA}$ for electrons from beauty is required to be less than approximately 0.85.
The measurements also appear to exclude pure QCD Bremsstrahlung in one calculation~\cite{Djordjevic:2005db}, but
as shown in the inset this calculation was already excluded by the total non-photonic electron $R_{AA}$.  Two other scenarios consistent with the total non-photonic
electron $R_{AA}$ are also consistent with the exclusion region.
Further progress is necessary to narrow this exclusion region and directly 
measure the $R_{AA}$ of charm and beauty separately.
  
\section{Outlook}
Charm and beauty have a wealth of decay modes; there are four pages
of decay modes of the $D$ and 10 pages of decay modes of the $B$ in the
Particle Data Book.  Not all of these are easily accessible in heavy ion collisions.
The most easily accessible modes, 
with large branching fractions uniformly across charm and beauty,
are the semi-leptonic decay modes, each of which is approximately 10\% per
leptonic species per hadron (though there are some perturbations for the
various charm states).  PHENIX's silicon upgrades mostly focus on these semi-leptonic
decay modes, with electrons at mid-rapidity with the VTX upgrade 
and muons at forward rapidity with the FVTX upgrade.
Charmed hadrons in general have shorter lifetimes than hadrons
containing beauty, so beauty can be distinguished from charm
by looking at displacements of the leptons from the primary vertex.
Projections indicate that the PHENIX detectors can
make precise measurements with this technique once installed.
Further distinguishing power can be obtained by combining the
correlation techniques of the previous section with displaced secondary
vertexing.  Such techniques combine information on the mass and 
lifetime of the parent hadron, and are standard in experiments
at other colliders.  The promising beginnings of such a study
using the VTX upgrade were shown in a poster from PHENIX~\cite{Lebedev}.  
For $B$ identification alone, the cleanest channel
is in the decay $B \rightarrow J/\Psi$, since a displaced J/$\Psi$ can only come from 
$B$ decays.  
STAR measurements of $J/\Psi$-hadron correlations indicate
that, at high $p_{T}$, (13 $\pm$ 5)\% of $J/\Psi$ come from $B$ decay~\cite{Abelev:2009qa}, implying that a meaningful sample
of such decays will be available once vertex detectors
are installed and RHIC II luminosities achieved.

The STAR Heavy Flavor Tracker, with its extremely precise
and thin inner layers, is designed specifically to allow
direct reconstruction, with low background,
of charmed hadrons from their hadronic decay modes, 
from rather low $p_{T}$ up to 10 GeV/c.  
This allows one to investigate the chemistry of charm fragmentation,
and search for its modification in heavy ion collisions, by reconstructing
and comparing a number of different charmed hadrons from $D^{0}$ to 
$\Lambda_{c}$.  For flow studies, total charm cross sections, and correlations,
the loss of information through the undetected neutrino in semi-leptonic
decays causes ambiguities.  These ambiguities prevent precise measurements in the hydrodynamic regime of parent $D$ $p_{T}$ below approximately 2 GeV/c, and can 
complicate precise interpretation of suppression patterns at higher $p_{T}$.  
Direct reconstruction of charmed hadrons using the Heavy Flavor Tracker
removes these ambiguities.  

Detectors at the LHC have been
built from the start with open heavy flavor in mind.  The landscape at the LHC
will likely be rather different than at RHIC.  Heavy flavor production will be quite a bit
more copious, since production cross sections for hadrons
containing heavier partons increase much 
more strongly with increasing $\sqrt{s}$ than those containing lighter partons.
This makes open heavy flavor, possibly including beauty,
 a promising tool for studying the chemistry
and flow of the bulk reacton zone.
At the same time, the $p_{T}$ scale will shift upwards, so that for studies
related to parton energy loss the charm quark will mostly act as a well-resolved
light quark.      
  
In summary, the study of open heavy flavor in heavy ion collisions is
off to a promising start, with a number of surprising hints that
to date have not been fully explained theoretically.  Currently these measurements
are limited, both qualitatively in their need to invoke assumptions, models, and
extrapolations, and quantitatively in their statistical resolving power.  
At RHIC, the first limitation
will soon be addressed via vertex detectors in both STAR and PHENIX,
and the second via the luminosity upgrade.   The landscape at the LHC remains to be seen.



\end{document}